\documentclass{aa}
\usepackage{epsfig}

\newcommand{\apjm}[2]{ApJ, #1, #2}

\newcommand{\aeta}[2]{A\&A, #1, #2}

\newcommand{\nh}{N$_{\rm H}$}

\newcommand{\Msol}{M$_{\odot}$ }

\newcommand{\rx}{\object{RX\,J0925.7-4758}}
\begin{document}
\title{XMM-Newton observations of MR Vel/\rx}

      \author{
       C. Motch 
      \inst{1}
      \and
      H. Bearda
      \inst{2,3}
      \and
      C. Neiner
      \inst{4}
        }
   
   \offprints{C. Motch, motch@astro.u-strasbg.fr}

   \institute{
              Observatoire Astronomique, UA 1280 CNRS, 11 rue de l'Universit\'e,
              F-67000 Strasbourg, France
              \and 
              Astronomical Institute, Utrecht University, P.O. Box 80000,
	      NL-3508 TA Utrecht, The Netherlands  
	      \and 
	      Terschelling 120, 3524 AZ Utrecht, The Netherlands
	      \and 	      
              Sterrenkundig Instituut Anton Pannekoek, 
              Kruislaan 403,
              NL-1098 SJ Amsterdam, The Netherlands
              }
              
   \date{Received ; accepted}

    \abstract{We report on XMM-Newton observations of the galactic supersoft X-ray source \rx . The
    RGS spectrum exhibits a wealth of spectral features from iron and oxygen. XMM-Newton data confirm
    the finding of previous Chandra HETGS/MEG observations that NLTE models of hot white dwarf
    atmospheres fail to represent the complex spectrum. There are clear evidences for P Cygni profiles
    with wind velocities of up to 2000 km\,s$^{-1}$. Small flux variations with time scales larger
    than 1000\,s are present. The strongest power is at $\sim$ 0.21\,d, a period close to that seen in
    V band optical light curves. A detailed analysis of the associated changes in the RGS and EPIC pn
    spectra hint at a mostly grey mechanism suggesting a variation of the visibility of the white
    dwarf due to occulting material in the accretion disk. Finally, we detect radial velocity changes
    of 173 $\pm$ 47 km\,s$^{-1}$ between two RGS observations obtained half an orbital cycle apart.
    The amplitude of the RGS velocity shift is consistent with that of the optical \ion{He}{ii}
    $\lambda$ 4686 and thus supports the idea that most of the \ion{He}{ii} optical line emission
    arises from the accretion disk.}

      
    \maketitle
%

\section{Introduction}

Supersoft X-ray sources have been discovered by the Einstein satellite and extensively studied by 
ROSAT. They usually consist of a close binary system in which the accreting component is a very
hot white dwarf responsible for the soft thermal-like radiation (van den Heuvel et al.
\cite{vdh}). At the low spectral resolution of the ROSAT PSPC, the energy distributions look like
black bodies with temperatures in the range of 20 to 70 eV. Higher resolution spectra with ASCA
(e.g. Ebisawa et al. \cite{ebisawa96}) started to reveal complex structures later emphasized by
Chandra and XMM-Newton grating spectroscopy (Paerels et al. \cite{pae01}, Bearda et al.
\cite{bea02}). Fitting model atmospheres of hot white dwarfs usually imply bolometric
luminosities below but close to Eddington. Stable or quasi stable hydrogen burning at the surface
of the white dwarf can adequately explain most of the observational features of supersoft X-ray
sources (van den Heuvel et al. \cite{vdh}). The high accretion rates required by stable
thermonuclear burning ($\dot{\rm M}$ $\sim$ 10$^{-7}$ \Msol yr$^{-1}$, Iben  \cite{iben82}) can be
obtained through thermally unstable Roche lobe overflow when the mass donor star is more massive
than the white dwarf. However, this picture  is probably not valid for many sources since short
periods systems cannot accommodate a massive mass-donor star. In these cases, strong
irradiation by the very luminous central source could play a major role in driving the mass
transfer mechanism (Van Teeseling \& King \cite{tee98}). 

Because stable burning allows matter to pile up at the surface of the white
dwarf, contrary to novae where most matter is expelled in the runaway
event, supersoft sources are considered as possible SN Ia progenitors. A sizeable
fraction of SN Ia could occur in these systems when the white dwarf approaches
the Chandrasekhar limit (Rappaport et al. \cite{rap94}).

We report in this paper on a long XMM-Newton observation of \rx \ involving the
EPIC pn, RGS and OM instruments. \rx \ is one of the few galactic supersoft source
known (Motch et al. \cite{mhp94}). It is identified with a V $\sim$ 17.2 reddened
object exhibiting Balmer and high excitation lines of \ion{He}{ii}, \ion{O}{vi} and
\ion{C}{iv}. The relatively high interstellar absorption towards the source (\nh \ 
$\sim$ 10$^{22}$ cm$^{-2}$) only allows observation in the highest energy range
(above $\sim$ 0.4 keV). Among supersoft sources, \rx \ is unusual by its relatively
high X-ray temperature and long orbital period of 4.0287 d (Schmidtke \&  Cowley
\cite{schmi01}). Previous X-ray observations have been reported from a number of
satellites (ASCA; Ebisawa et al. \cite{ebisawa01}, BeppoSAX; Hartmann et al.
\cite{hh99} and recently using the HETG on board Chandra; Bearda et al.
\cite{bea02}). 

\section{XMM-Newton and optical observations}

The XMM-Newton satellite is described in Jansen et al. (\cite{jansen}). XMM-Newton
observations took place in late December 2000 during two consecutive satellite orbits,
thus covering two phase intervals of the binary period half a cycle apart. The original
scheduling was done using the ephemeris from Schmidtke et al. (\cite{schmi00}) which
assumed a 3.83\,d period and was planned to cover photometric phases 0.0 (minimum) and
0.5. With the revised period of 4.0287\,d (Schmidtke \&  Cowley \cite{schmi01}) the
actual range of phase covered is slightly shifted towards earlier times. The journal of
X-ray observations is listed in Table \ref{jlog}. We also organized quasi-simultaneous
optical observations using the Dutch telescope on La Silla during one night. In December,
\rx \ was only observable in the last part of the night just after the end of the first
XMM-Newton observation. 

The EPIC background remained nominal during the two observations apart for the last
2000\,s of the second one. EPIC pn data were acquired in the large window mode in
order to minimize pile-up problems from this relatively bright source (up to 7
cts/s in EPIC pn). The MOS was left in imaging mode but the source count rate is
too high for proper spectral analysis due to severe pile up effects. The OM was
used in normal imaging mode and because of the presence of bright stars in the
field of view, the UVW1 filter was used. RGS data had to be completely
reprocessed from the ODF late October 2001 when good calibrations at high energies
eventually became available. All data have been analyzed using SAS version
20010917\_1110. 

Optical observations were carried out using the standard CCD camera at the Dutch
telescope located on ESO La Silla, Chile. V and B filters were alternated with exposure
times of 3 and 5 minutes in the V and B filters respectively. Standard MIDAS procedures
were used to correct images for flat-field and bias.  Instrumental magnitudes were
computed using the Sextractor package (Bertin \& Arnouts \cite{be96}) and linked to
comparison star 2 in Motch et al. (\cite{mhp94}).

\begin{table}
\caption{Journal of Observations}
\begin{tabular}{ll}
\hline
\multicolumn{2}{l}{Rev 187: 2000-12-16 11:01 - 2000-12-17 04:05} \\
\multicolumn{2}{l}{Photometric phase : 0.82 - 0.98} \\
\hline
Instrument & Exposure time (s) \\
EPIC MOS 1+2 & 58400 \\
EPIC pn & 56700 \\
RGS 1+2 & 61200 \\
\hline
\multicolumn{2}{l}{Rev 188: 2000-12-18 10:51 - 2000-12-19 04:06} \\
\multicolumn{2}{l}{Photometric phase : 0.31 - 0.47} \\
\hline
Instrument & Exposure time (s) \\
EPIC MOS 1+2 & 59000 \\
EPIC pn & 57300 \\
RGS 1+2 & 61600 \\
\hline
\end{tabular}
\label{jlog}
\end{table}

\section{Source variability}

\begin{figure}[]

\psfig{figure=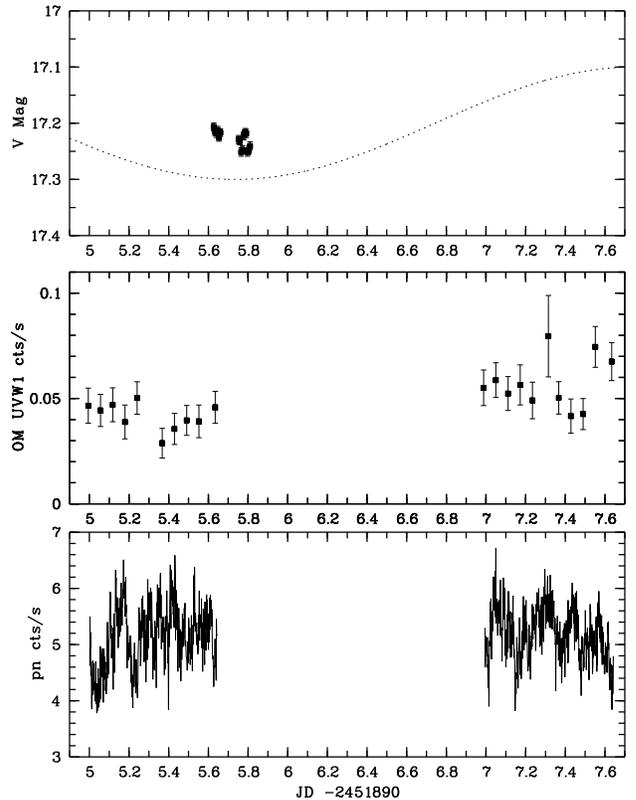,width=9cm,bbllx=1.5cm,bburx=19cm,bblly=2cm,bbury=24.0cm,clip=true}

\caption[]{Simultaneous EPIC pn (0.5-1.0 keV, 100\,s bins), OM and V band Dutch data. The dotted line in the
upper panel represents the mean photometric V light curve taken from Schmidtke et al. (\cite{schmi00})
assuming the revised orbital period of 4.0287 d}

\label{simdata}
\end{figure}

Fig. \ref{simdata} shows the simultaneous EPIC pn and OM UVW1 and the contemporaneous
V band light curve obtained at the Dutch telescope.  The EPIC pn light curve displays
significant variability at the level of ten percent on various time scales. Some
kind of oscillation with a period of $\sim$ 0.2d is clearly visible as well
as faster variability. Not surprisingly considering the high interstellar absorption
towards the source (\nh \ $\sim$ 10$^{22}$ cm$^{-2}$) OM UVW1 data are of much lower
statistical quality, each point representing an integration over 5000 seconds. There
is no compelling evidence for any strong correlations between X-ray and UV flux
variations.  During the XMM-Newton observations, the mean V and B magnitudes are
similar to those usually observed ($<V>$  = 17.23 and $<B-V>$ = 1.88) indicating that
\rx \ was not in an especially bright or low optical state. The V level is only $\sim$
0.07 mag above the mean value from Schmidtke et al. (\cite{schmi00}), consistent with
the observed cycle to cycle intrinsic variability. 

\begin{figure}[]

\psfig{figure=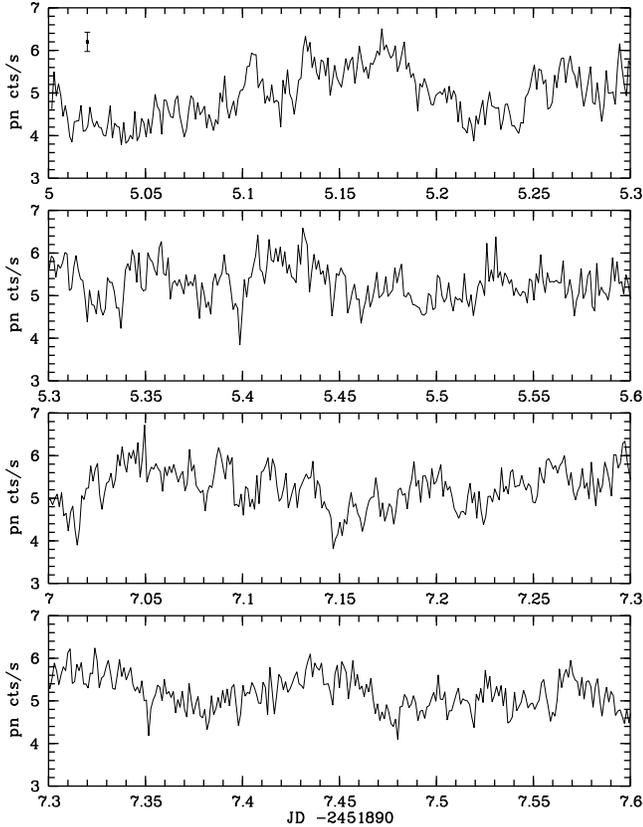,width=9cm,bbllx=2.5cm,bburx=19.5cm,bblly=2cm,bbury=24.0cm,clip=true}

\caption[]{Details of the EPIC pn light curve(0.5-1.0 keV, 100 sec bins).  A typical
error bar is shown in the upper panel. Variations on time scales larger than  $\sim$
1000 sec are clearly seen in the power spectrum}

\label{pnlc}
\end{figure}

\begin{figure}[]

\psfig{figure=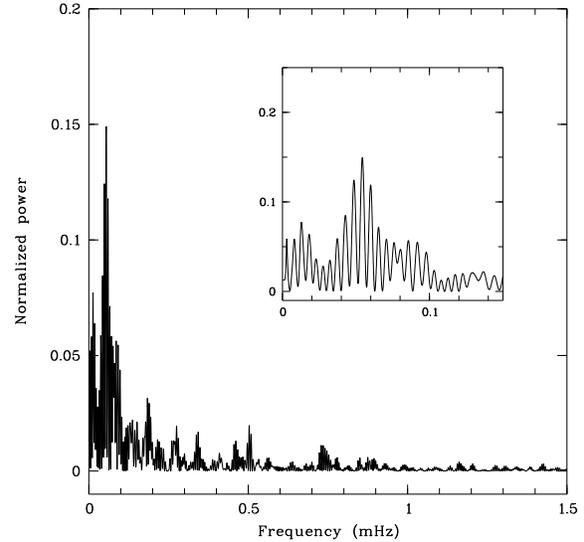,angle=-90,width=8cm,bbllx=0.5cm,bburx=20cm,bblly=1.5cm,bbury=21.5cm,clip=true}

\caption[]{Power spectrum of the combined pn EPIC light curve of the two observations (0.5-1.0
keV). The highest peak is at P $\sim$ 0.21d, a period close to that reported for the optical
quasi-oscillations by Schmidtke et al. \cite{schmi00}}

\label{plotpnpower}
\end{figure}

We searched for coherent oscillations in each individual observation as well as in the combined
data sets using the least square power spectrum method outlined in Lomb (\cite{lomb76}) and Scargle
(\cite{scargle82}). Light curves with bin sizes of 10 and 100 seconds (0.5-1.0 keV) do not show any
statistically significant pulsations for periods in the range of 30 to 1000\,s. There is however a
clear power excess at low frequencies on time scales larger than 1000 seconds which can be seen in
the detailed light curve (Fig. \ref{pnlc}) and in the power spectrum of the combined EPIC pn
observations (see Fig. \ref{plotpnpower}). The highest peak in the periodogram of each individual
observation as well as in that of the merged EPIC pn light curves is at P $\sim$ 0.21\,d. Its
formal statistical significance is high and the modulation is well seen in Fig. \ref{simdata}. Best
periods in revolutions 187 and 188 are P = 0.2152 $\pm$ 0.0059 d and P = 0.1977 $\pm$ 0.0082 d
respectively and are therefore consistent with a constant value. The periodogram of the merged EPIC
pn data has its highest peak at P = 0.2130 $\pm$ 0.009 d. Observations over many more cycles would
be needed to be able to measure the stability of this X-ray modulation and especially compare it
with that of the 0.23\,d quasi-periodicity detected in the optical by Schmidtke et al.
(\cite{schmi00}). The mean full amplitude of the best fit sine wave to the 0.2\,d modulation is
10.8 \%. 

\section{RGS Spectroscopy}

\subsection{Overall spectrum}

\begin{figure*}[ht]

\psfig{figure=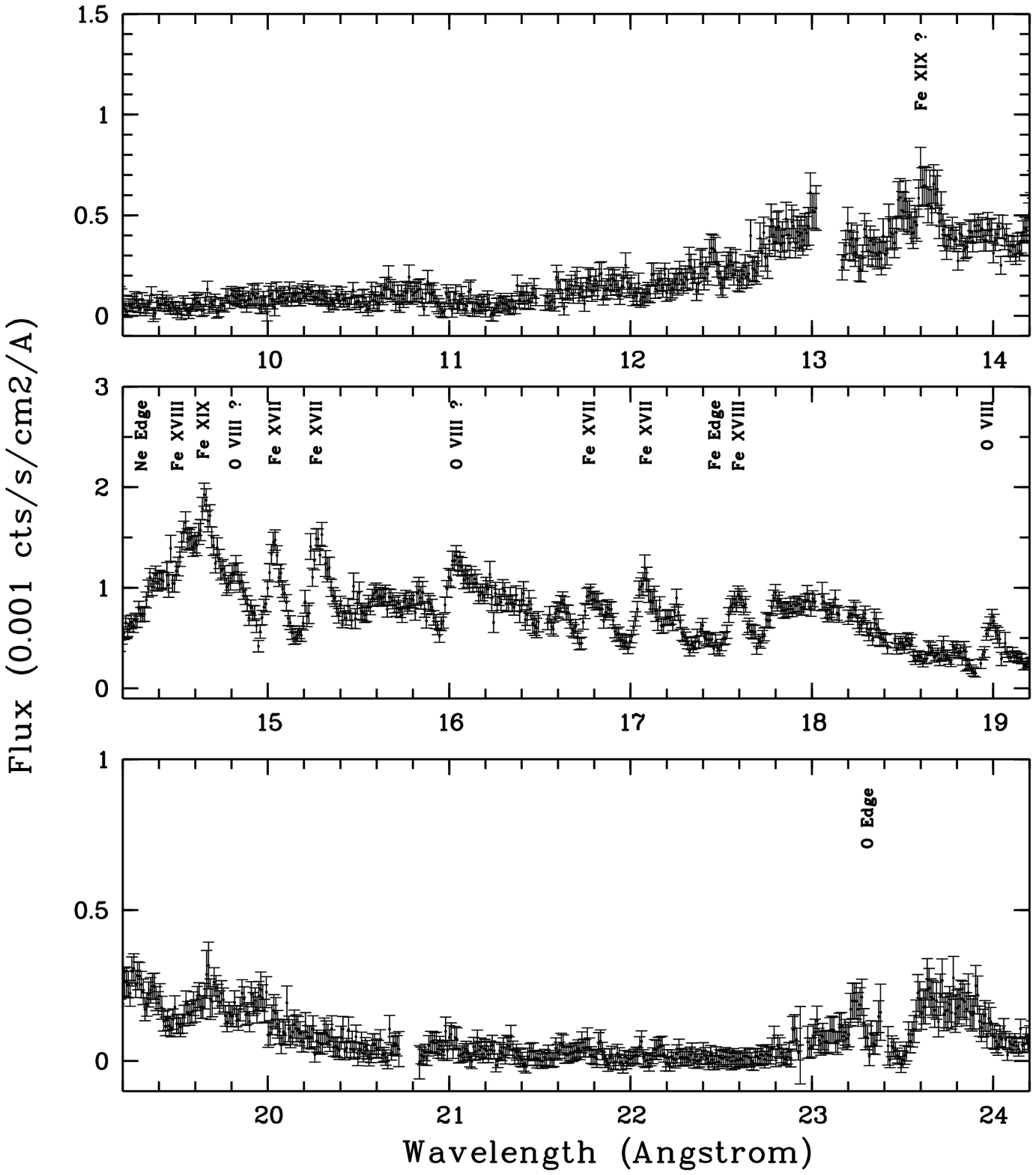,width=17cm,height=20cm,bbllx=2cm,bburx=19cm,bblly=1cm,bbury=21.0cm,clip=true}

\caption[]{The RGS spectrum of \rx \ summed over the two observations and the two RGS
instruments and corrected for effective area.}

\label{rgsspec}

\end{figure*}

\begin{figure}[]

\psfig{figure=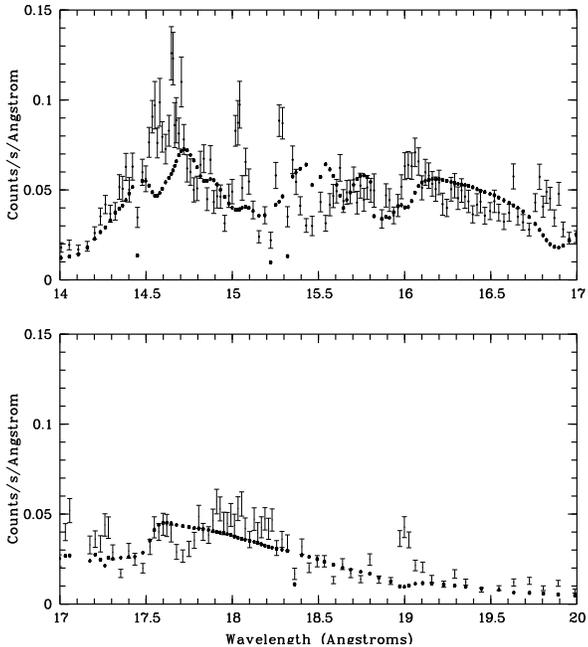,width=8cm,bbllx=2cm,bburx=19cm,bblly=6cm,bbury=26.0cm,clip=true}

\caption[]{The RGS - 1 spectrum of \rx \ summed over the first observation. The 'best' fit hot
white dwarf NLTE model atmosphere of Bearda et al. (\cite{bea02}) with \nh\ = 1.28 10$^{22}$
cm$^{-2}$ is shown for comparison. }

\label{plothenkmodel}

\end{figure}

Thanks to the high throughput of the RGS and long total exposure time ($\sim$ 120 ks), the
spectrum of \rx \ summed over the two observations and the two RGS reveals an unprecedented
number of spectral features compared to the Chandra HETG spectrum discussed in Bearda et al.
(\cite{bea02}). We show in Fig. \ref{rgsspec} the total XMM-Newton spectrum together with
tentative line identifications. This spectrum was accumulated using SAS task rgsfluxer and the
corresponding reprocessed extractions and response matrices computed for each individual
instrument and observation. Spectral bins with large associated errors have been removed from
the plot for clarity. All features identified in the Chandra spectrum can be seen in the RGS
data. The spectrum is clearly not black body like and exhibits a large number of high excitation
iron lines, mostly seen in emission as well as lines of the \ion{O}{viii} Lyman series. The
conclusions of Bearda et al. (\cite{bea02}) that the structures seen in the Chandra spectrum
cannot be reproduced by LTE or NLTE model atmospheres can only be emphasized by our RGS data.
Fig. \ref{plothenkmodel} shows the 'best' fit NLTE white dwarf model atmosphere of Bearda et al.
(\cite{bea02}) which has log g = 9 and T = 8.9 10$^{5}$ K. The intensity of the RGS spectrum in
the O edge region suggests an absorption of \nh\ = 1.50 10$^{22}$ cm$^{-2}$ slightly higher than
that assumed by Bearda et al. (\cite{bea02}) for HETGS/MEG data (\nh\  = 1.0 10$^{22}$ cm$^{-2}$).
However, ignoring this long wavelength region, a better fit is obtained in the 17 to 20 \AA \
range with \nh\  = 1.28 10$^{22}$ cm$^{-2}$. Overall, the picture is similar to that drawn from
Chandra data, namely that although the shape of the energy distribution seems to be grossly
described by the model, strong emission features such as the \ion{O}{viii} lines or the
\ion{Fe}{xvii} lines at $\lambda$ 15.015 \AA\ and $\lambda$ 15.262 \AA \ are not present in the
model atmosphere. We did not investigate further spectral modelling, leaving this study open for
future work. The lack of X-ray flux in the RGS and EPIC below $\sim$ 0.4 keV argues for a strong
interstellar absorption consistent with that seen at optical wavelengths (\nh\ $\sim$
10$^{22}$cm$^{-2}$). Interstellar O edge is well marked at 23.3 \AA \ and there are clear
evidences for the presence of Fe-L and Ne edges at 17.54 \AA \ and 14.30 \AA \ respectively,
although, part of these structures could also be due to intrinsic spectral features from the hot
atmosphere. 

In the absence of proper modelling of the emission spectrum it is impossible to derive fundamental
parameters such as temperature, density, gravity and luminosity. Also, since most of the lines are
suspected to be blended, line profile distortions due to bulk motion for instance are difficult to
analyse in details. There seems however to be a clean case of a P Cygni profile in the Lyman
$\alpha$ \ion{O}{viii} line at 18.97 \AA , which is relatively free of  contaminating lines from
other species. As noted by Bearda et al. (\cite{bea02}), other Fe lines have P Cygni - like
profiles as well as the probable identification of  Lyman $\beta$ \ion{O}{viii} line at 16.00 \AA
\ which is however, suspected of being blended with Fe lines. The blue edge of the mean Ly
$\alpha$ \ion{O}{viii} P Cygni profile shown in Fig. \ref{oviii} is at velocities close to 2000
km\,s$^{-1}$ slightly larger than the 1500 km\,s$^{-1}$ derived by Bearda et al. (\cite{bea02}),
but much less than that of the transient collimated wind or jet discovered by Motch (\cite{m98}).

\subsection{Variability with orbital phases and flux level}

The RGS 1 and 2 summed spectra for revolutions 187 and 188 are identical within the error bars.
Over the 13 to 20 \AA \ range the source appears slightly brighter by $\sim$ 12 $\pm$ 4 \% in
the last observation. However, the flux increase seems to occur over the whole energy
range and the difference RGS spectrum does not show marked features which would give hints
towards a change in the physical conditions of the source or of its environment.

\begin{figure}[]

\psfig{figure=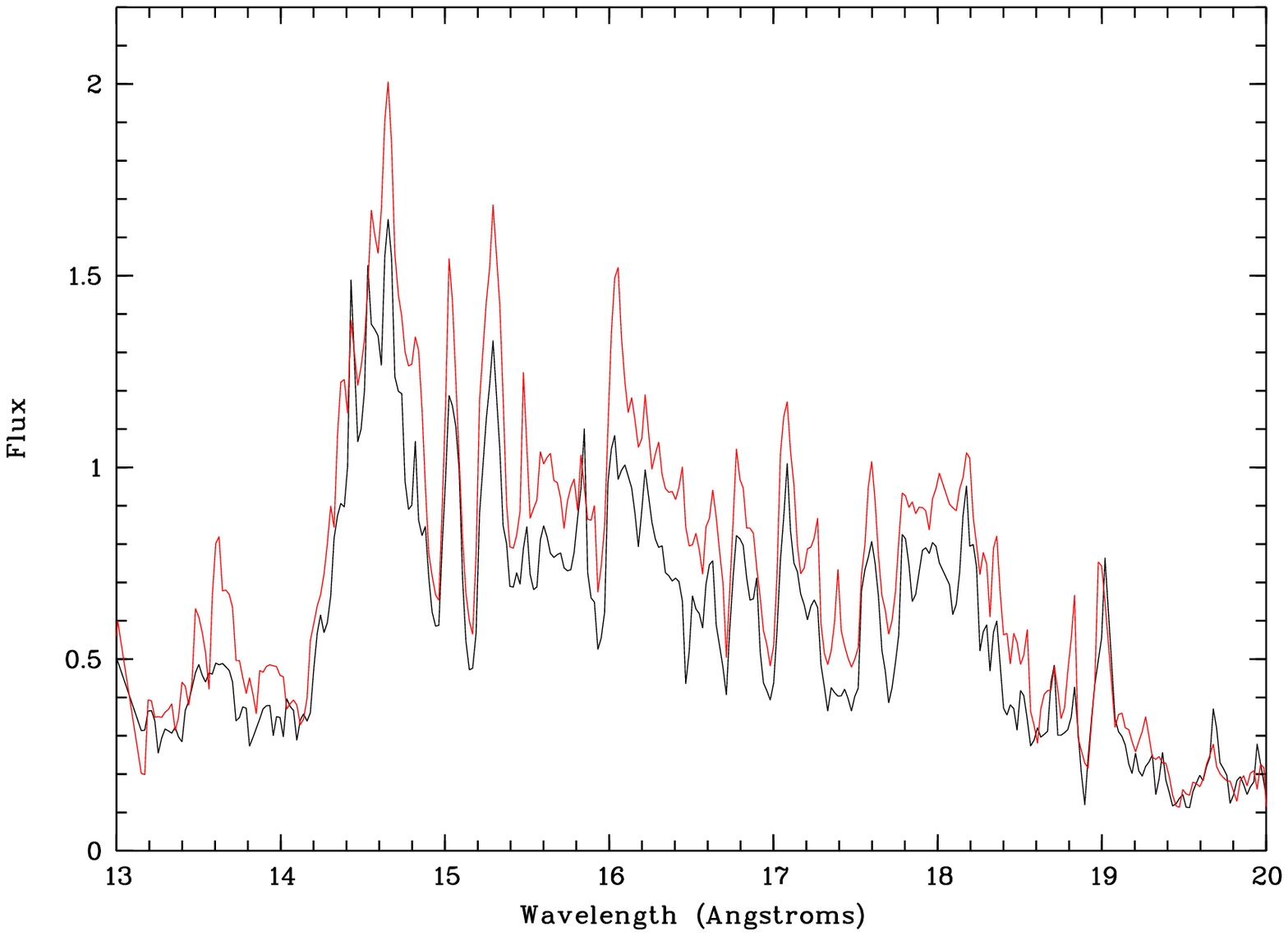,width=8cm,clip=true}

\caption[]{RGS 1 + RGS 2 spectra accumulated for the low and high flux states}

\label{rgslowhigh}

\end{figure}

\begin{figure}[]

\psfig{figure=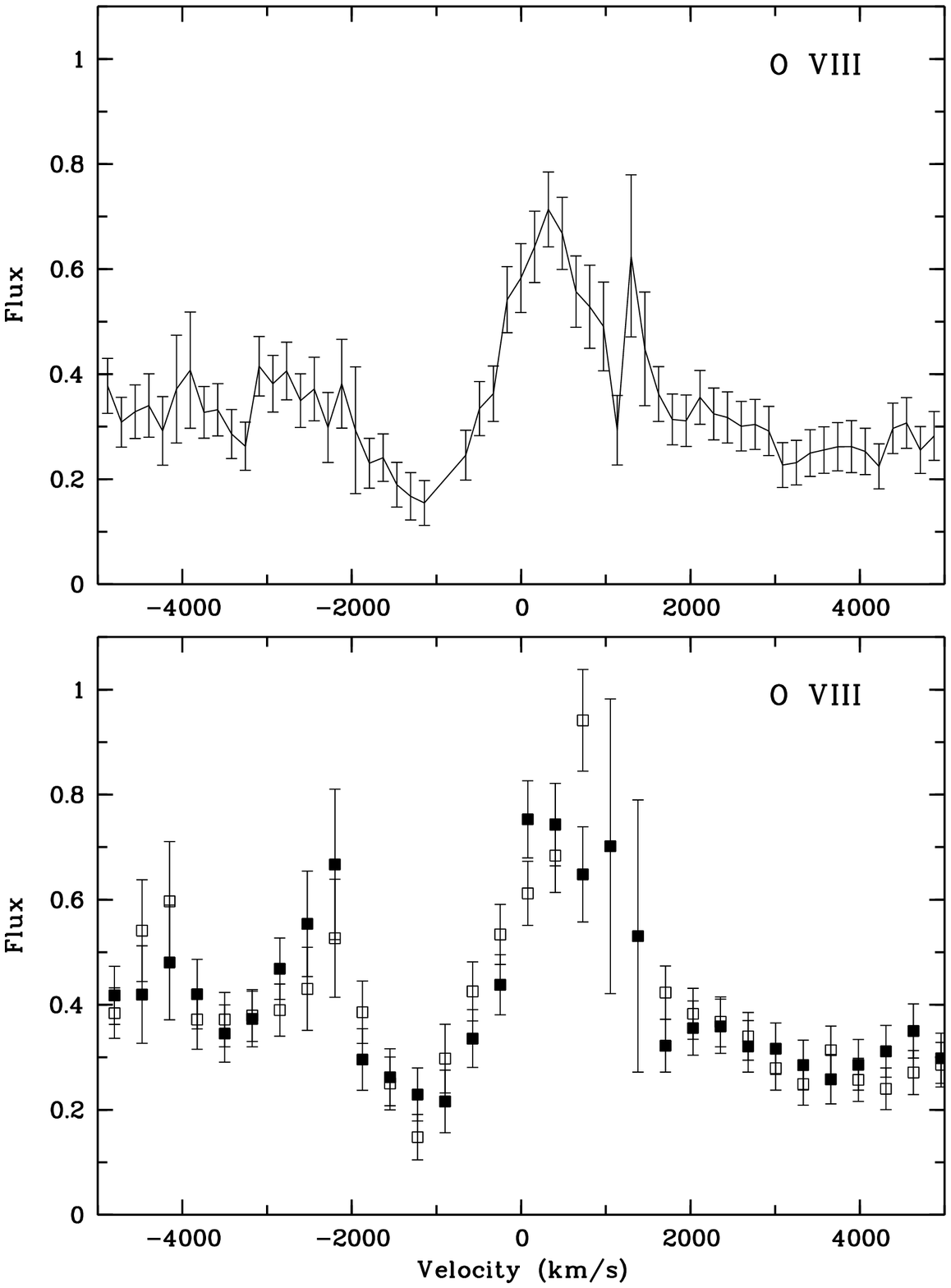,width=8cm,angle=0,,bbllx=2.0cm,bburx=20cm,bblly=1.0cm,bbury=25cm,clip=true}  

\caption[]{Top: \ion{O}{viii} Ly $\alpha$ P Cygni profile summing all RGS data. Bottom : 
low flux (open squares) and high flux (filled squares) \ion{O}{viii} Ly $\alpha$ P Cygni
profiles. Low flux data have been multiplied by 1.23 to fit the mean high flux level.}

\label{oviii}

\end{figure}

The variability seen in the EPIC light curve on time scales larger than 1000\,s is also detected in
the RGS light curves. We thus tried to see whether marked differences could be found in the line
intensities of the low and high flux states in order to shed light on the mechanism giving rise to the
small flux variations. For that purpose, we created RGS light curves by selecting events in the area
of the $\beta$ / x-disp RGS plane where the source lies (see den Herder et al., \cite{denherder}
for a description of the RGS on board XMM-Newton). For each observation, a summed RGS 1 and 2 light
curve was created with 100 sec bins. Two time intervals series corresponding to count rates below and
above 0.73 cts/s and 0.61 cts/s for the first and second observations respectively were used to
accumulate low and high state RGS spectra with similar exposure times. The last part of the second
observation, contaminated by high background has been discarded. SAS task rgsfluxer was then used to
add together low and high state spectra of different instruments and observations. Overall the high to
low count rate ratio is 1.23. The two spectra  shown in Fig. \ref{rgslowhigh} are very similar. The
only noticeable differences are a possible higher emissivity at $\sim$ 13.62 \AA \ (\ion{Fe}{xix}?),
16.05 (\ion{Fe}{xviii},\ion{O}{viii} ?) and 17.40 \AA \ and lower emissivity at $\sim$ 14.44 \AA ,
14.52 \AA (\ion{Fe}{xviii} ?) in the high flux spectrum. The Lyman $\alpha$ \ion{O}{viii} P Cygni
profile, once corrected for the different flux levels shows some evidence for a flux dependence.
Although the terminal velocity remains close to the $\sim$ 2000 km\,s$^{-1}$ seen in the average
profile (Fig. \ref{oviii}), the low velocity part of the profile seems more absorbed (or the emission
less strong) at low flux.  The two RGS individually give similar pictures, albeit with a lower
significance. 

\begin{figure}[tbp]

\psfig{figure=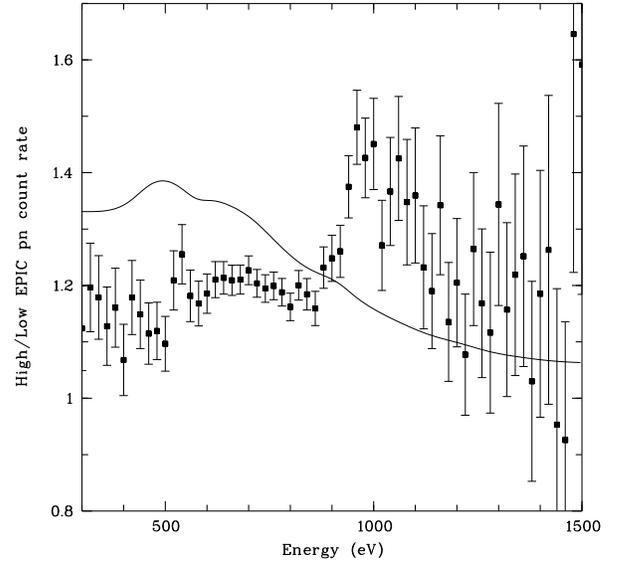,width=8cm,angle=-90,clip=true}

\caption[]{High to low EPIC pn count ratio plotted versus energy. The solid line represents the
spectral changes expected if intensity variations were only due to photoelectric absorption by
cold plasma.}

\label{plotpnratio}

\end{figure}

We applied the same analysis to EPIC pn spectra, using this time EPIC pn count rates to define three
flux levels (high, low and medium) with equal corresponding exposure times. The mean EPIC pn count
rate of \rx \ is $\sim$ 6 cts/s, well below the limit of 12 cts/s set for pile-up in the large window
mode. In order to minimize any residual effect due to pile-up we restricted our analysis to pn pattern
= 0 events. Fig. \ref{plotpnratio} shows the ratio of the high to low EPIC pn spectra plotted for the
merged revolutions 187 and 188 data. Although the NLTE model atmosphere of Bearda et al. (\cite{bea02})
does not satisfactorily fit the EPIC pn spectrum, it provides a better representation than black body
spectra and can be used to estimate the expected distortion of the energy distribution if it was only
due to photoelectric absorption. In the 0.5-1.0 keV range the high to low ratio of 1.277 can be
mimicked by a change from \nh\ = 1.485 10$^{22}$ to 1.540 10$^{22}$ cm$^{-2}$, keeping all other
parameters constant. The almost 'grey' flux variation from 400 to 900 eV is clearly at variance with
what would be expected from photoelectric absorption by cold material with changing column density.
The high to low flux ratio is significantly larger above $\sim$ 940 eV. Unfortunately RGS data are not
sensitive enough to confirm the EPIC pn behavior. In the absence of overall spectral model, no clear
explanation exists yet for this high energy bump. Remaining pile-up effects could account for part or
all of the increase. However, results from the diagnostics SAS 5.3 task epatplot suggest that pile-up
is not strong enough to account for all of the observed effect.

\section{Radial velocities}

\begin{figure*}[tbp]

\begin{tabular}{cc}
\psfig{figure=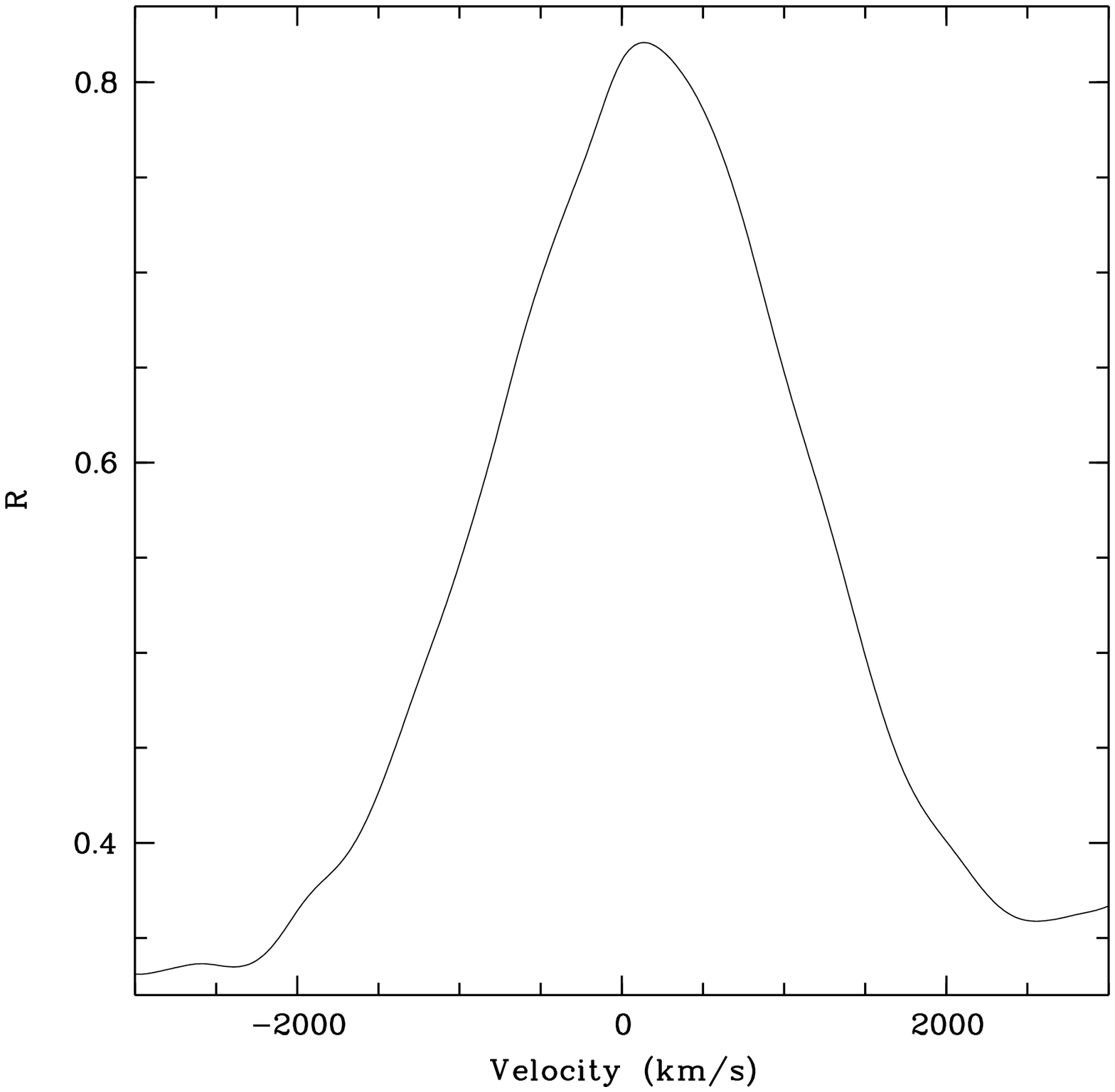,width=8cm,angle=-90,,bbllx=1.0cm,bburx=21cm,bblly=1.0cm,bbury=22cm,clip=true} & 
\psfig{figure=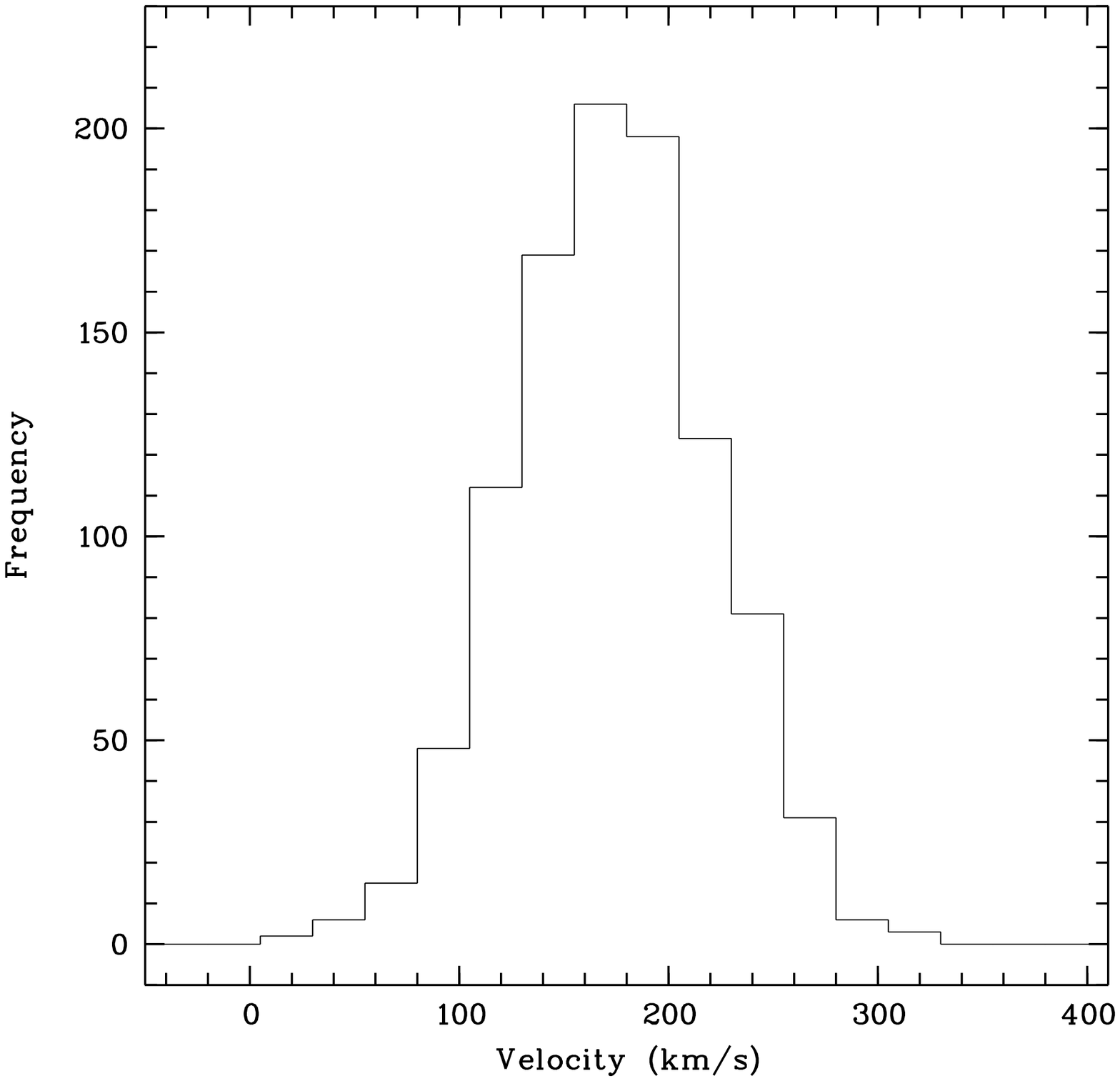,width=8cm,angle=-90,,bbllx=1.0cm,bburx=21cm,bblly=1.0cm,bbury=22cm,clip=true}
\end{tabular}

\caption[]{Left: Cross-correlation curve of the 14.18\,\AA \ to 18.55\,\AA\  RGS 1 + 2 spectra
obtained in revolutions 187 and 188. Right: Distribution of Gaussian fit centers to the
cross-correlation peaks of the same randomized spectra repeated a thousand times. The mean
velocity is 173 $\pm$ 47 km\,s$^{-1}$.}

\label{velocity}

\end{figure*}

Although we do not have yet any spectral model adequate for representing RGS data, the weakness of
the observed spectral changes with either orbital phase or X-ray flux level offers the possibility
to search for radial velocity variations, ignoring thus the complications of the unknown
underlying physics and taking advantage of the wealth of relatively sharp spectral features. The
lack of identified unblended lines from the stellar atmosphere does not allow to compute absolute
velocities and furthermore the RGS wavelength calibration may not be known with enough accuracy.
However, comparison of the two observations obtained two days apart may allow to reveal radial
velocity changes with orbital phase. For that purpose we applied a cross correlation algorithm,
frequently used in optical spectroscopy and described in details in Tonry \& Davis
(\cite{tondav79}). Spectra were created using SAS task rgsfluxer for each instrument and
observation. Their wavelength step of 0.0103 \AA \ samples well the  instrument FWHM resolution
which is $\sim$ 0.07 \AA \ ($\sim$ 1200 km\,s$^{-1}$) in the 14 - 19 \AA \ range where most of the
flux is present. After removing wavelength bins with high associated errors the log $\lambda$
spectra were linearly interpolated over a velocity grid with constant steps of 15 km\,s$^{-1}$.
Absorption edges produced by cold material have much less amplitude and structures than those of
emission lines from the hot atmosphere. Therefore, these features which may be partly fixed in
velocity should have little influence on the amplitude of global Doppler shifts. In the wavelength
range 14.18 \AA \ to 18.55 \AA \ the peak resulting from the cross-correlation of the merged RGS 1
and 2 spectra from revolutions 187 and 188  has a Gaussian shape and its center is displaced by
$\sim$ 170 km\,s$^{-1}$ (see Fig. \ref{velocity}, left panel).

In order to estimate the significance of this velocity shift we generated artificial RGS spectra adding to
the observed data a random value normally distributed with a sigma equal to the assigned error of the
wavelength bin. In this manner we can realistically account for the varying statistical quality over the
spectrum. These artificial spectra were cross-correlated applying the same procedure as the real ones and
the resulting peak was fitted by a Gaussian. This process was typically repeated between 1,000 and 5,000
times. The distribution of the Gaussian centroids is clearly shifted  whatever the wavelength interval used
between 14.18 \AA \ and 18.55 \AA \ and independently of the instrument considered, R1 alone, R2 alone or R1
+ R2 merged (see Fig. \ref{velocity}, right panel). We list in Table \ref{radvel} the results of the
Monte Carlo error calculation for different configurations. 

\begin{table}
\caption{Radial velocity shifts between revolutions 187 and 188}
\begin{tabular}{ccc}
Detector & $\lambda$ range (\AA ) & Velocity (km\,s$^{-1}$)\\
\hline
RGS 1 + RGS 2 & 14.18 -- 18.85 & 173 $\pm$ 47 \\
RGS 1 + RGS 2 & 14.94 -- 15.43 & 290 $\pm$ 93 \\
RGS 1 + RGS 2 & 16.45 -- 17.72 & 164 $\pm$ 71 \\
RGS 1         & 14.18 -- 18.85 & 194 $\pm$ 80 \\
RGS 2         & 14.18 -- 18.85 & 159 $\pm$ 67 \\
\end{tabular}
\label{radvel}
\end{table}

Obviously, the most accurate determination is obtained by using the largest wavelength range and
the two RGS detectors. However, the measured displacements are all consistent especially those
derived from the completely independent RGS 1 and RGS 2 instruments.  The revised ephemeris of
Schmidtke \& Cowley (\cite{schmi01}) and phase of minimum velocity read from their Fig. 4 predict
for the \ion{He}{ii} $\lambda$ 4686 an average velocity difference of $\sim$ 120 to 140
km\,s$^{-1}$ between observations in revolutions 187 ($\Phi_{Phot}$ = 0.818-0.977) and 188
($\Phi_{Phot}$ = 0.312-0.473). This compares well with the best RGS value of 173 $\pm$ 47
km\,s$^{-1}$ and gives further evidence that the velocity shift seen in the RGS is real. The sigma
of the Gaussian cross correlation peak indicates an average FWHM width of $\sim$ 3600 km\,s$^{-1}$.
This width is significantly larger than the intrinsic resolution of the RGS and suggests that line
blending and/or wind effects such as seen in the Lyman $\alpha$  \ion{O}{viii} P Cygni profile
affect most of the spectrum. 

In order to check these results we fitted with xspec Gaussian profiles on the top of a flat continuum
to two of the largest and apparently weakly blended lines at 15.015 \AA (\ion{Fe}{xvii}) and close to 17.6
\AA (\ion{Fe}{xviii}). We used the raw background corrected spectra and their corresponding response matrix
files to fit the two RGS simultaneously. This method is in principle the most correct one since profile
distortions are explicitely taken into account. Note however, that as long as spectral features do not have
orbital phase dependent profiles, as it seems to be the case here, the cross correlation method applied to
spectra generated by the SAS task rgsfluxer should yield correct results. Xspec fits indicate a mean
displacement of 173 $\pm$ 73\,km\,s$^{-1}$ comparable to that resulting from the cross
correlation method. Since we only consider here two lines the significance of the velocity shift is
expectedly smaller than when using all  spectral features together. However, the fact that the shifts
derived from xspec are comparable to those computed from the spectra created by rgsfluxer is an independent
confirmation of the reality of the velocity difference betweeen the two RGS observations.

\section{Discussion and Conclusions}

XMM-Newton RGS spectra have revealed a wealth of spectral features, some of them being
apparently resolved by the RGS. Emission and also maybe absorption lines due to various iron
ions are seen as well as Lyman $\alpha$ \ion{O}{viii}. The RGS spectra confirm all the features
seen in the Chandra HETGS by Bearda et al. (\cite{bea02}) and illustrate further the difficulty
to correctly model the emission of hot white dwarf atmospheres. Clearly, more sophisticated models
including wind effects are required. 

The X-ray spectrum of \rx \ display little changes with both X-ray flux level and orbital phase.
At the start of the first observation, the X-ray source is receding with maximum positive
\ion{He}{ii} velocity while two days later the source moves towards us with maximum negative
\ion{He}{ii} velocities. This means that the geometrical configuration of the system is not very
different between the two observations. In the absence of wake material, the column density due to
the wind of the mass donor star does not change drastically and this is at first order consistent
with the lack of large phase dependence of the RGS spectrum.

Very few lines show evidences for possible intensity related variations and the EPIC pn clearly
indicates that absorption by cold material is not the mechanism responsible for the X-ray
modulations. Interestingly, the 0.21\,d period for X-ray oscillations is very close to the 0.23\,d
preferred time scale reported by Schmidtke et al. (\cite{schmi00}) in the optical light curve. The
full amplitude of sine fit to the soft X-ray flux modulation at 0.21\,d is $\sim$ 11\% while that
of the optical V band modulation is between two and three times less ($\sim$ 4\% Schmidtke et al.
\cite{schmi00}). Since we do not have simultaneous optical V band and X-ray observations we do not
know the exact flux ratio nor phasing of the modulations. OM data are not of high enough S/N to
reveal a modulation at the level of a few percents. Observations at the Dutch telescope were too
short to constrain the amplitude of a contemporaneous 0.2\,d optical periodicity. However, such an
optical to X-ray amplitude ratio is expected when optical emission is dominated by X-ray heating of
the companion star or of the accretion disk (see e.g. Van Paradijs \& Mc. Clintock \cite{vm94},
Matsuoka et al. \cite{ma84}). A change in the visible emitting area of the accreting white dwarf by
the internal rim of the accretion disc could thus explain both the X-ray and optical 0.2\,d
modulation. Simultaneous X-ray and optical photometric observations may reveal beating effects at
the 4\,d orbital period and could thus constrain the amount of optical light resulting from X-ray
heating of the mass donor star and/or of the accretion disk bulge.

The RGS spectra also confirm the existence of a strong wind clearly affecting the Lyman $\alpha$
\ion{O}{viii} emission line and probably many other lines. The wind velocity of $\sim$ 2000
km\,s$^{-1}$ is more than two times slower than the velocity of the transient jet detected by Motch
(\cite{m98}) at 5200 km\,s$^{-1}$ or of that of the candidate receding $\lambda$ 6680.3 \AA \
component proposed by Schmidtke et al. (\cite{schmi00}) at 5350 km\,s$^{-1}$. In the absence of
realistic model atmosphere including wind effects, it is difficult to strictly rule out the presence
of weak satellite lines at $\pm$ 5200 km\,s$^{-1}$ in the RGS spectrum. The autocorrelation of the RGS
spectra do show some signal at similar velocity shifts. However, a more likely explanation is the
presence of several iron lines with the adequate spacing in wavelength.

Amazingly, the amplitude and direction of the RGS radial velocity changes with orbital phase (173
$\pm$ 47 km\,s$^{-1}$) is fully compatible with that expected from the \ion{He}{ii} $\lambda$ 4686
optical line. This observation thus supports the idea that in this system at least, the
\ion{He}{ii} $\lambda$ 4686 velocity represents the orbital motion of the X-ray source.  This may
be one of the first observation measuring orbital Doppler shifts in high energy spectral features
from an X-ray source. Our coverage in orbital phase is not dense enough to derive the amplitude
and phasing of the X-ray lines with respect to the optical \ion{He}{ii} line. As for the H$\alpha$
line, some of the \ion{He}{ii} emission may arise from the heated hemisphere of the companion star
or from the bulge of the accretion disc. This additional component may reduce the amplitude of the
\ion{He}{ii} velocity change with respect to K$_{\rm X}$ and add some phase lag. A long RGS
observation of \rx \ may thus be able to measure with relatively high accuracy the true orbital
velocity of the accreting object. A reliable K$_{\rm X}$ estimate could help constraining the mass
ratio and check whether the high mass transfer in \rx \ can be driven by thermally unstable Roche
lobe overflow. 

\begin{acknowledgements}

We thank E. Rol for carrying the optical observations at the Dutch telescope on La Silla and E.
Janot-Pacheco for helping in the organization of the observations. We are also grateful to the
referee for suggesting several improvements and to M.W. Pakull for useful discussions.

\end{acknowledgements}

\end{document}